\UseRawInputEncoding

\documentclass[reprint, twocolumn, showkeys]{revtex4}

\usepackage{graphicx}
\usepackage{dcolumn}
\usepackage{bm}
\usepackage{color}
\usepackage{amsmath,amsthm}
\usepackage{amssymb,bm}
\usepackage{mathrsfs}
\usepackage{comment}
\usepackage{ulem}

\graphicspath{%
    {converted_graphics/}
    {/}
}
\begin{document}

\title{Semiclassical Truncated-Wigner-Approximation Theory of Molecular Exciton-Polariton Dynamics in Optical Cavities}
\author{Nguyen Thanh Phuc}
\email{nthanhphuc@moleng.kyoto-u.ac.jp}
\affiliation{Department of Molecular Engineering, Graduate School of Engineering, Kyoto University, Kyoto 615-8510, Japan}
\affiliation{PRESTO, Japan Science and Technology Agency (JST), Kawaguchi, Japan}

\begin{abstract}
Molecular exciton polaritons are hybrid states resulting from the strong coupling of molecular electronic excitations with an optical cavity mode, presenting a promising approach for controlling photophysical and photochemical properties in molecular systems. 
In this study, we develop a semiclassical theory for molecular exciton-polariton dynamics using the truncated Wigner approximation (TWA) to explore the collective behavior of molecular electronic excited states under strong light-matter coupling.
Our approach expands the previously developed TWA theory for molecular vibration-polariton dynamics (J. Chem. Theory Comput. 2024, 20, 3019--3027) by incorporating semiclassical treatment of quantum coherence between ground and excited molecular states.
We initially apply the TWA theory to a simplified system of molecules modeled as two-level (spin-1/2) systems, omitting vibronic coupling. 
The semiclassical results derived from applying the TWA to single-spin operators demonstrate excellent agreement with full quantum dynamic simulations in systems with a sufficiently large number of molecules.
Lastly, the TWA theory is extended to incorporate molecular vibronic coupling, revealing the dynamic polaron decoupling effect, where quantum coherence between molecular excitations is preserved under strong light-matter coupling.
\end{abstract}


\maketitle

\section{Introduction}
Molecular exciton polaritons represent a fascinating class of hybrid light-matter states that emerge when molecular electronic excitations are strongly coupled with an optical cavity mode. 
This interaction forms new quasiparticles, which combine the properties of both exciton (molecular excited states) and photons (cavity modes). 
The field of exciton-polaritons has garnered considerable interest due to the potential for controlling photophysical and photochemical properties of molecular systems in novel ways~\cite{Ebbesen16, Vidal21}. 
By tuning the light-matter coupling strength, researchers can significantly alter the energy landscape~\cite{Hutchison12} and dynamic behavior of molecules, offering new possibilities for applications in enegy conversion, molecular electronics, and catalysis~\cite{Hertzog19}. 

One of the most exciting aspects of molecular exciton polaritons is their ability to modify the excited-state dynamics of molecules within optical cavities. 
Numerous experimental studies have demonstrated their impact on a wide range of processes, such as altering reaction kinetics, enhancing or suppressing energy transfer pathways~\cite{Munkhbat18}, and enabling the inversion of excited-state populations~\cite{Eizner19}. 
For instance, strong light-matter coupling has been used to enhance the conductivity of organic materials~\cite{Orgiu15}, modulate singlet fission rates~\cite{Takahashi19}, and control selective emission pathways through polariton funneling~\cite{Satapathy21}. 
These phenomena present intriguing opportunities for the design of next-generation materials with tailored photonic and electronic properties. 

From a theoretical perspective, strong light-matter interactions give rise to several collective and quantum effects that are not present in conventional molecular systems. 
These include collective enhancement in reaction rates via superreaction mechanisms~\cite{Phuc21}, Bose-enhanced energy transfer in polariton condensates~\cite{Phuc22}, cavity-mediated superconductivity~\cite{Schlawin19}, and photon-coupled chiral-induced spin selectivity~\cite{Phuc23}.
However, modeling these systems is computationally challenging due to the sheer number of molecular degrees of freedom involved and the complexity of their interactions with the cavity mode. 
Full quantum dynamic simulations become intractable as the system size increases, especially when dealing with molecules interacting collectively with a single cavity mode. 

Given the computational limitations of fully quantum approaches, it is necessary to develop efficient and scalable methods that can accurately capture the essential quantum features of exciton-polariton dynamics while allowing for the simulation of large molecular ensembles. 
The semiclassical truncated Wigner approximation (TWA) offers a promising solution to this problem~\cite{Moyal49, Hillery84, Polkovnikov10}. 
The TWA provides a way to approximate quantum dynamics by evolving classical trajectories in phase space while still retaining key quantum mechanical effects, such as quantum fluctuations and coherence. 
In this work, we build upon the previous TWA framework for molecular vibration-polariton dynamics~\cite{Phuc24}, extending it to include quantum coherence between electronic ground and excited states. 
To model a discrete set of $\mathcal{N}$ electronic eigenstates in a molecule, we apply the generalized discrete TWA, where the dynamic observables are expressed as a set of $\mathcal{N}^2$ Hermitian operators. 
These operators correspond to the generalized Gell-Mann matrices supplemented by the identity matrix~\cite{Zhu19}. 
For molecules with only a single excited state ($\mathcal{N}=2$), the Gell-Mann matrices simplify to the three Pauli matrices, commonly used for spin-1/2 systems~\cite{Schachenmayer15}. 

To validate the TWA in this context, we first apply the theory to a simplified model of molecules treated as two-level systems (i.e., spin-1/2 systems), omitting vibronic coupling. 
This approach allows us to focus on the purely electronic degrees of freedom and their interaction with the cavity mode. 
By comparing the semiclassical results to full quantum dynamic simulations, we demonstrate that the TWA yields accurate results for large molecular systems, even in the ultrastrong coupling regime, where the light-matter interaction strength is comparable to the energy scales of the molecules and cavity mode. 
This is due to the enhanced mean-field behavior and reduced influences of quantum correlation and nonlinearity in large systems. 
However, when compared to pure mean-field theory results, it becomes clear that omitting the sampling from the Wigner distribution significantly reduces the accuracy of the predictions. 
Finally, by incorporating nuclear degrees of freedom and including vibronic coupling, we extend the TWA approach to examine the decay of quantum coherence between electronic excitations in different molecules. 
Vibronic interactions introduce additional complexity by coupling electronic excitations to nuclear motions, leading to decoherence and energy dissipation.
However, we observe that strong light-mater coupling suppresses the decay of quantum coherence, consistent with predictions based on the dynamic polaron decoupling effect~\cite{Spano15, Herrera16, Phuc19, Takahashi20, Phuc21}. 

\section{TWA theory of molecular exciton-polariton dynamics}
We consider a system consisting of $N$ identical molecules, each with electronic excitations that are strongly coupled to a single-mode optical cavity. 
The cavity mode has a frequency $\omega_\text{c}$, and the light-matter interaction is described using the Coulomb gauge Hamiltonian~\cite{Tannoudji-book, Mandal23}
\begin{align}
\hat{H}=&\sum_{n=1}^N
\left[\sum_j \frac{\left(\hat{\mathbf{P}}_{n,j}-q_j\hat{\mathbf{A}}\right)^2}{2M_j}
+\sum_k \frac{\left(\hat{\mathbf{p}}_{n,k}+e\hat{\mathbf{A}}\right)^2}{2m_\text{e}}
\right] \nonumber\\
&+\sum_{n=1}^N \left[\hat{V}^{(n)}\left(\left\{\hat{\mathbf{R}}_{n,j}\hat{\mathbf{r}}_{n,k}\right\}\right)
+\frac{1}{2}\sum_{l\not=n} \hat{V}_\text{int}^{(nl)}\right]
+\hbar\omega_\text{c}\hat{a}^\dagger\hat{a},
\label{eq: minimal coupling Hamiltonian}
\end{align}
where $\hat{\mathbf{R}}_{n,j},\hat{\mathbf{P}}_{n,j}$ and $\hat{\mathbf{r}}_{n,j},\hat{\mathbf{p}}_{n,j}$ represent the position and momentum operators for the $j$-th nucleus and the $k$-th electron in the $n$-th molecule, with corresponding charges $q_j$ and $-e$ and masses $M_j$ and $m_\text{e}$, respectively. 
The terms $\hat{V}^{(n)}$ and $\hat{V}_\text{int}^{(nl)}$ represent the intra- and inter-molecular interactions, respectively. 
The cavity field is quantized using the annihilation and creation operators $\hat{a}$ and $\hat{a}^\dagger$, and the vector potential operator $\hat{\mathbf{A}}$ for the cavity mode is expressed as
\begin{align}
\hat{\mathbf{A}}=\mathbf{A}_0\left(\hat{a}+\hat{a}^\dagger\right),
\end{align}
where $\mathbf{A}_0$ is the amplitude of the vector potential in the cavity's vacuum field.

To simplify the Hamiltonian, we eliminate the $\hat{\mathbf{A}}^2$ term using the Bogoliubov transformation~\cite{Liberato17}
\begin{align}
\hat{a}=\left(\cosh r\right) \hat{c}-\left(\sinh r\right) \hat{c}^\dagger,
\end{align}
with $e^r=\sqrt{\tilde{\omega}_\text{c}/\omega_\text{c}}$, where $\tilde{\omega}_\text{c}=\sqrt{\omega_\text{c}^2+2\alpha^2}$, and $\alpha$ is defined as
\begin{align}
\alpha=\sqrt{2\omega_\text{c}C/\hbar}
\label{eq: alpha}
\end{align} 
with $C$ given by
\begin{align}
C=NA_0^2\left(\sum_j \frac{q_j^2}{2M_j}+N_\text{e}\frac{e^2}{2m_\text{e}}\right)
\simeq \frac{NN_\text{e}e^2A_0^2}{2 m_\mathrm{e}}.
\end{align}
Here, $N_\text{e}$ is the number of electrons in each molecule, and in the approximation, we neglect the contribution from the nuclei due to their significantly larger masses compared to electrons.
After applying the Bogoliubov transformation, the total Hamiltonian is rewritten as
\begin{align}
\hat{H}=&\sum_{n=1}^N
\left[\sum_j \frac{\hat{\mathbf{P}}_{n,j}^2}{2M_j}
+\sum_k \frac{\hat{\mathbf{p}}_{n,k}^2}{2m_\text{e}}
\right] \nonumber\\
&+\sum_{n=1}^N \left[\hat{V}^{(n)}\left(\left\{\hat{\mathbf{R}}_{n,j}\hat{\mathbf{r}}_{n,k}\right\}\right)
+\frac{1}{2}\sum_{l\not=n} \hat{V}_\text{int}^{(nl)}\right]
+\hbar\tilde{\omega}_\text{c}\hat{c}^\dagger\hat{c}\nonumber\\
&+\frac{e}{m_\text{e}}\sqrt{\frac{\omega_\text{c}}{\tilde{\omega}_\text{c}}}\left(\hat{c}+\hat{c}^\dagger\right)
\sum_{n=1}^N \mathbf{A}_0\cdot\left(\sum_k \hat{\mathbf{p}}_{n,k}\right),
\label{eq: Hamiltonian after Bogoliubov transformation}
\end{align}
where the interaction of the cavity field with the nuclei has been neglected due to the much larger nuclear masses and the off-resonant frequency.
The annihilation and creation operators $\hat{c}$ and $\hat{c}^\dagger$ continue to satisfy the commutation relation for bosons: $\left[\hat{c},\hat{c}^\dagger\right]=1$.

For simplicity, we neglect intermolecular interactions, though these can be included in a generalized version of the theory.
Using the Born-Oppenheimer approximation, the electronic Hamiltonian for each molecule is diagonalized for a fixed nuclear configuration $\left\{\mathbf{R}_{n,j}\right\}$,  yielding a set of electronic eigenstates $|u_\nu^{(n)}\left(\left\{\mathbf{R}_{n,j}\right\}\right)\rangle$ and corresponding eigenvalues $\epsilon_\nu^{(n)}\left(\left\{\mathbf{R}_{n,j}\right\}\right)$:
\begin{align}
\hat{H}_\text{el}^{(n)}|u_\nu^{(n)}\rangle=\epsilon_\nu^{(n)}|u_\nu^{(n)}\rangle.
\end{align}
In the case where the system does not enter the ultrastrong coupling regime (where the light-matter interaction strength is comparable to the molecular excitation energies), the analysis can be restricted to a limited subspace of $\mathcal{N}$ electronic eigenstates. 
In this subspace, the total Hamiltonian becomes
\begin{align}
\hat{H}=&\sum_{n=1}^N
\left[\sum_j \frac{\hat{\mathbf{P}}_{n,j}^2}{2M_j}
+\sum_{\nu=1}^\mathcal{N} \epsilon_\nu^{(n)}|u_\nu^{(n)}\rangle\langle u_\nu^{(n)}|\right]
+\hbar\tilde{\omega}_\text{c}\hat{c}^\dagger\hat{c}\nonumber\\
&+\left(\hat{c}+\hat{c}^\dagger\right)
\sum_{n=1}^N\sum_{\nu<\mu}^\mathcal{N}\left[g_{\nu\mu}^{(n)}|u_\nu^{(n)}\rangle\langle u_\mu^{(n)}|+\text{h.c.}\right],
\end{align}
where h.c. stands for Hermitian conjugate, and the light-matter coupling strengths $g_{\nu\mu}^{(n)}$ are defined as
\begin{align}
g_{\nu\mu}^{(n)}\left(\left\{\mathbf{R}_{n,j}\right\}\right)=\frac{e}{m_\text{e}}\sqrt{\frac{\omega_\text{c}}{\tilde{\omega}_\text{c}}}
\Bigg\langle u_\nu^{(n)}\Bigg|\mathbf{A}_0\cdot\left(\sum_k \hat{\mathbf{p}}_{n,k}\right)\Bigg|u_\mu^{(n)}\Bigg\rangle.
\end{align}
The diagonal matrix elements of the momentum operator $\hat{\mathbf{p}}_{n,k}$ vanish, as $\hat{\mathbf{p}}_{n,k}=(i m_\text{e}/\hbar)\left[\hat{H}_\text{el}^{(n)},\hat{\mathbf{r}}_{n,k}\right]$, leading to
\begin{align}
\langle u_\nu^{(n)}|\hat{\mathbf{p}}_{n,k}|u_\mu^{(n)}\rangle=&\frac{i m_\text{e}}{\hbar}\left(\epsilon_\nu^{(n)}-\epsilon_\mu^{(n)}\right)\left\langle u_\nu^{(n)}\Big|\hat{\mathbf{r}}_{n,k}\Big|u_\mu^{(n)}\right\rangle.
\end{align}

For an $\mathcal{N}\times\mathcal{N}$ Hilbert subspace, we introduce a set of $\mathcal{N}^2$ Hermitian operators $\hat{\Lambda}_\tau^{(n)} (\tau=1,\cdots,\mathcal{N}^2, n=1,\cdots,N)$, using the generalized Gell-Mann matrices and the identity matrix $\hat{\text{I}}^{(n)}$ as a complete basis for the electronic degrees of freedom in each molecule~\cite{Zhu19}:
\begin{align}
\hat{\Lambda}_{\tau=(\mu-2)(\mu-1)/2+\nu}^{(n)}=&
\frac{1}{\sqrt{2}}(|u_\nu^{(n)}\rangle\langle u_\mu^{(n)}|+\text{h.c.}) \nonumber\\
&\text{ for } 1\leq \nu<\mu\leq \mathcal{N},
\end{align}
\begin{align}
\hat{\Lambda}_{\tau=(\mu-2)(\mu-1)/2+\nu+\mathcal{N}(\mathcal{N}-1)/2}^{(n)}=&
\frac{1}{i\sqrt{2}}(|u_\nu^{(n)}\rangle\langle u_\mu^{(n)}|-\text{h.c.}) \nonumber\\
&\text{ for } 1\leq \nu<\mu\leq \mathcal{N},
\end{align}
\begin{align}
\hat{\Lambda}_{\tau=\mathcal{N}(\mathcal{N}-1)+\mu}^{(n)}=&
\frac{1}{\sqrt{\mu(\mu+1)}}\Bigg(\sum_{\nu=1}^\mu |u_\nu^{(n)}\rangle\langle u_\nu^{(n)}|\nonumber\\
&-\mu|u_{\mu+1}^{(n)}\rangle\langle u_{\mu+1}^{(n)}|\Bigg) \nonumber\\
&\text{ for } 1\leq \mu\leq \mathcal{N}-1,
\end{align}
\begin{align}
\hat{\Lambda}_{\tau=\mathcal{N}^2}^{(n)}=&
\sqrt{\frac{1}{\mathcal{N}}}\hat{\text{I}}^{(n)}.
\end{align}
These operators are orthonormal, satisfying the relation $\text{Tr}\left\{\hat{\Lambda}_\tau^{(n)}\hat{\Lambda}_{\chi}^{(n)}\right\}=\delta_{\tau\chi}$, and they follow the commutation relation
\begin{align}
\left[\hat{\Lambda}_\tau^{(n)},\hat{\Lambda}_\chi^{(n)}\right]=i\sum_\xi s_{\tau\chi\xi}\hat{\Lambda}_\xi^{(n)},
\end{align}
where the structure constants $s_{\tau\chi\xi}$ are defined by $i s_{\tau\chi\xi}=\text{Tr}\left\{\hat{\Lambda}_\xi^{(n)}\left[\hat{\Lambda}_\tau^{(n)},\hat{\Lambda}_\chi^{(n)}\right]\right\}$.
Using these operators, the electronic degrees of freedom in each molecule can be fully described.
The inverse relations for the projection operators are given by
\begin{align}
|u_\nu^{(n)}\rangle\langle u_\mu^{(n)}|=&\frac{1}{\sqrt{2}}\left(\hat{\Lambda}_\tau^{(n)}+i\hat{\Lambda}_{\tau+\frac{\mathcal{N}(\mathcal{N}+1)}{2}}^{(n)}\right),\\
|u_\mu^{(n)}\rangle\langle u_\nu^{(n)}|=&\frac{1}{\sqrt{2}}\left(\hat{\Lambda}_\tau^{(n)}-i\hat{\Lambda}_{\tau+\frac{\mathcal{N}(\mathcal{N}+1)}{2}}^{(n)}\right)
\end{align}
for $1\leq \nu<\mu\leq \mathcal{N}$, with $\tau=(\mu-2)(\mu-1)/2+\nu$.
The diagonal projection operators can be written as
\begin{align}
|u_\nu^{(n)}\rangle\langle u_\nu^{(n)}|=&\frac{\hat{\text{I}}^{(n)}}{\mathcal{N}}
+\sum_{\mu=1}^{\mathcal{N}-\nu} \frac{1}{\sqrt{(\mathcal{N}-\mu)(\mathcal{N}-\mu+1)}}\hat{\Lambda}_{\mathcal{N}^2-\mu}^{(n)}\nonumber\\
&-\sqrt{\frac{\nu-1}{\nu}}\hat{\Lambda}_{\mathcal{N}^2-\mathcal{N}+\nu-1}^{(n)}
\end{align}
for $1\leq\nu\leq\mathcal{N}-1$, and
\begin{align}
|u_\mathcal{N}^{(n)}\rangle\langle u_\mathcal{N}^{(n)}|=&\frac{\hat{\text{I}}^{(n)}}{\mathcal{N}}
-\sqrt{\frac{\mathcal{N}-1}{\mathcal{N}}}\hat{\Lambda}_{\mathcal{N}^2-1}^{(n)}.
\end{align}
These relations enable the total Hamiltonian to be expressed in terms of the $\hat{\Lambda}_\tau^{(n)}$ operators:
\begin{align}
\hat{H}=&\hbar\tilde{\omega}_\text{c}\hat{c}^\dagger\hat{c}
+\sum_{n=1}^N\Bigg\{\sum_j \frac{\hat{\mathbf{P}}_{n,j}^2}{2M_j}
+\frac{\hat{\text{I}}^{(n)}}{\mathcal{N}}\sum_{\nu=1}^\mathcal{N}\epsilon_\nu^{(n)}\left(\left\{\hat{\mathbf{R}}_{n,j}\right\}\right)\nonumber\\
&+\sum_{\nu=1}^{\mathcal{N}-1}\hat{\Lambda}_{\mathcal{N}^2-\mathcal{N}+\nu}^{(n)}f_\nu^{(n)}\left(\left\{\hat{\mathbf{R}}_{n,j}\right\}\right)\nonumber\\
&+\sqrt{2}\left(\hat{c}+\hat{c}^\dagger\right)\sum_{\nu<\mu}^\mathcal{N}\Bigg[\hat{\Lambda}_{(\mu-2)(\mu-1)/2+\nu}^{(n)}\text{Re}g_{\nu\mu}^{(n)}\left(\left\{\hat{\mathbf{R}}_{n,j}\right\}\right)\nonumber\\
&-\hat{\Lambda}_{(\mu-2)(\mu-1)/2+\nu+\mathcal{N}(\mathcal{N}-1)/2}^{(n)}\text{Im}g_{\nu\mu}^{(n)}\left(\left\{\hat{\mathbf{R}}_{n,j}\right\}\right)\Bigg]\Bigg\}.
\end{align}
The functions $f_\nu^{(n)}$ are defined as
\begin{align}
f_\nu^{(n)}=\frac{1}{\sqrt{\nu(\nu+1)}}\sum_{\mu=1}^\nu\epsilon_\mu^{(n)}-\sqrt{\frac{\nu}{\nu+1}}\epsilon_{\nu+1}^{(n)}.
\end{align}
The Heisenberg equations of motions for the quantum operators are
\begin{align}
\frac{\text{d}\hat{\mathbf{R}}_{n,j}}{\text{d}t}=\frac{\hat{\mathbf{P}}_{n,j}}{M_j},
\end{align}
\begin{align}
\frac{\text{d}\hat{\mathbf{P}}_{n,j}}{\text{d}t}=&-\frac{\hat{\text{I}}^{(n)}}{\mathcal{N}}\sum_{\nu=1}^\mathcal{N}\nabla_j\epsilon_\nu^{(n)}
-\sum_{\nu=1}^{\mathcal{N}-1}\hat{\Lambda}_{\mathcal{N}^2-\mathcal{N}+\nu}^{(n)}\nabla_j f_\nu^{(n)}\nonumber\\
&-\sqrt{2}\left(\hat{c}+\hat{c}^\dagger\right)\sum_{\nu<\mu}^\mathcal{N}\Bigg[\hat{\Lambda}_{(\mu-2)(\mu-1)/2+\nu}^{(n)}\text{Re}\nabla_j g_{\nu\mu}^{(n)}\nonumber\\
&-\hat{\Lambda}_{(\mu-2)(\mu-1)/2+\nu+\mathcal{N}(\mathcal{N}-1)/2}^{(n)}\text{Im}\nabla_j g_{\nu\mu}^{(n)}\Bigg],
\end{align}
\begin{align}
\frac{\text{d}\hat{c}}{\text{d}t}=&-i\tilde{\omega}_c\hat{c}
-\sqrt{2}\sum_{n=1}^N\sum_{\nu<\mu}^\mathcal{N}\Bigg[\hat{\Lambda}_{(\mu-2)(\mu-1)/2+\nu}^{(n)}\text{Re}g_{\nu\mu}^{(n)}\nonumber\\
&-\hat{\Lambda}_{(\mu-2)(\mu-1)/2+\nu+\mathcal{N}(\mathcal{N}-1)/2}^{(n)}\text{Im}g_{\nu\mu}^{(n)}\Bigg],
\end{align}
\begin{align}
\frac{\text{d}\hat{\Lambda}_\tau^{(n)}}{\text{d}t}=&-\frac{1}{\hbar}\sum_{\xi=1}^{\mathcal{N}^2}
\Bigg\{\sum_{\nu=1}^{\mathcal{N}-1}s_{(\mathcal{N}^2-\mathcal{N}+\nu)\tau\xi} f_\nu^{(n)}+\sqrt{2}\left(\hat{c}+\hat{c}^\dagger\right)\nonumber\\
&\sum_{\nu<\mu}^\mathcal{N}
\Bigg[s_{((\mu-2)(\mu-1)/2+\nu)\tau\xi}\text{Re}g_{\nu\mu}^{(n)}\nonumber\\
&-s_{((\mu-2)(\mu-1)/2+\nu+\mathcal{N}(\mathcal{N}-1)/2)\tau\xi}\text{Im}g_{\nu\mu}^{(n)}\Bigg]\Bigg\}.
\end{align}
Here, $\nabla_j$ represents the partial derivative with respect to the nuclear coordinates $\mathbf{R}_{n,j}$, and for simplicity, the explicit dependence of $\epsilon_\nu^{(n)}$, $f_\nu^{(n)}$, and $g_{\nu\mu}^{(n)}$ on the nuclear coordinates $\mathbf{R}_{n,j}$ has been omitted.

In the TWA framework, the expectation value of an arbitrary operator is approximated by averaging the corresponding Weyl symbol over the phase-space distribution, which is initially represented by the Wigner function~\cite{Moyal49, Hillery84, Polkovnikov10, Phuc24}. 
The equations of motion for the classical variables $\mathbf{R}_{n,j}$, $\mathbf{P}_{n,j}$, $c$, and $\Lambda_\tau^{(n)}$ are obtained by replacing the quantum operators with their classical counterparts in the Heisenberg equations of motion.
We assume that the photonic, nuclear, and electronic degrees of freedom are initially uncorrelated, so the total density operator factorizes as
\begin{align}
\hat{\rho}_\text{tot}(t=0)=\hat{\rho}_\text{ph}(t=0)\otimes \hat{\rho}_\text{nu}(t=0)\otimes \hat{\rho}_\text{el}(t=0).
\end{align} 
For the photonic and nuclear degrees of freedom, the initial Wigner functions are given by
\begin{align}
W_\text{ph}(c)=&\int\frac{\text{d}^2\eta}{\pi^2}
e^{\eta^*c-\eta c^*}
\text{Tr}\left\{\hat{\rho}_\text{ph}(t=0) e^{\eta\hat{c}^\dagger-\eta^*\hat{c}}\right\}
\end{align}
and
\begin{align}
W_\text{nu}(\mathbf{R},\mathbf{P})=&\hbar^D \int\frac{\text{d}^D\mathbf{u}}{(2\pi)^D} \int\frac{\text{d}^D\mathbf{v}}{(2\pi)^D} 
e^{i(\mathbf{u}\cdot\mathbf{P}+\mathbf{v}\cdot\mathbf{R})}\nonumber\\
&\text{Tr}\left\{\hat{\rho}_\text{nu}(t=0)e^{-i(\mathbf{u}\cdot\hat{\mathbf{P}}+\mathbf{v}\cdot\hat{\mathbf{R}})}\}\right\},
\end{align}
where $\mathbf{R}=\left\{\mathbf{R}_{n,j}\right\}$, $\mathbf{P}=\left\{\mathbf{P}_{n,j}\right\}$, $\mathbf{u}=\left\{\mathbf{u}_{n,j}\right\}$, and $\mathbf{v}=\left\{\mathbf{v}_{n,j}\right\}$ are the sets of coordinate and momentum variables, and $D$ is the total number of nuclear degrees of freedom.
For the electronic degrees of freedom, a discrete probability distribution is used~\cite{Schachenmayer15, Zhu19}. 
Specifically, each operator $\hat{\Lambda}_\tau^{(n)}$ can be decomposed into its eigenvectors $|\lambda_\tau^{(n)}\rangle$ with corresponding eigenvalues $\lambda_\tau^{(n)}$, such that $\hat{\Lambda}_\tau^{(n)}=\sum_{\lambda_\tau^{(n)}}\lambda_\tau^{(n)}|\lambda_\tau^{(n)}\rangle\langle\lambda_\tau^{(n)}|$.
In a projective measurement, the eigenvalues $\lambda_\tau^{(n)}$ represent the possible outcomes of measuring $\hat{\Lambda}_\tau^{(n)}$.
If there are no initial correlations between molecules for the electronic state, the initial density operator for the electronic degrees of freedom is a product state: $\hat{\rho}_\text{el}(t=0)=\prod_{n=1}^N \hat{\rho}_\text{el}^{(n)}(t=0)$, and the probability distribution for the classical variable $\Lambda_\tau^{(n)}$ is given by
\begin{align}
p_\tau^{(n)}\left(\Lambda_\tau^{(n)}=\lambda_\tau^{(n)}\right)=&
\text{Tr}\left\{\hat{\rho}_\text{el}^{(n)}(t=0)|\lambda_\tau^{(n)}\rangle\langle\lambda_\tau^{(n)}|\right\}.
\label{eq: discrete probability distribution}
\end{align}
Here, the possible values of the $\Lambda_\tau^{(n)}$ is limited to a set of discrete values $\lambda_\tau^{(n)}$.
The overall distribution factorizes for different variables both within the same molecule and between molecules. 
This means that the probability of a specific configuration of the set $\left\{\Lambda_\tau^{(n)}\right\}$ being a combination of the eigenvalues $\lambda_\tau^{(n)}$ is given by
\begin{align}
p\left(\left\{\Lambda_\tau^{(n)}=\lambda_\tau^{(n)}\right\}\right)=\prod_{n=1}^N\prod_{\tau=1}^{\mathcal{N}^2-1}
p_\tau^{(n)}\left(\Lambda_\tau^{(n)}=\lambda_\tau^{(n)}\right).
\end{align}
For diagonal initial states, all initial correlations between $\hat{\Lambda}_\tau^{(n)}$ operators can be perfectly reproduced by this discrete probability distribution~\cite{Zhu19}.
Even for non-diagonal pure states, it is possible to transform the state into a diagonal form via a local unitary transformation, allowing the TWA to effectively capture the initial electronic state.

\section{TWA dynamics of exciton polaritons}
For simplicity, in the following we consider a system where the dynamics are restricted to two electronic states: the ground state $|\text{g}\rangle$ and the first excited state $|\text{e}\rangle$, i.e., $\mathcal{N}=2$.
In this case, the three operators $\hat{\Lambda}_\tau (\tau=1,2,3)$ correspond to the Pauli matrices:
\begin{align}
\hat{\Lambda}_1
=&\frac{|\text{e}\rangle\langle\text{g}|+|\text{g}\rangle\langle\text{e}|}{\sqrt{2}}
=\frac{\hat{\sigma}_x}{\sqrt{2}}, \\
\hat{\Lambda}_2
=&\frac{|\text{g}\rangle\langle\text{e}|-|\text{e}\rangle\langle\text{g}|}{i\sqrt{2}}
=\frac{\hat{\sigma}_y}{\sqrt{2}},\\
\hat{\Lambda}_3
=&\frac{|\text{g}\rangle\langle\text{g}|-|\text{e}\rangle\langle\text{e}|}{\sqrt{2}}
=\frac{\hat{\sigma}_z}{\sqrt{2}}.
\end{align}
We first analyze the coupled dynamics of electronic and photonic degrees of freedom, neglecting both nuclear degrees of freedom and cavity loss, where the system exhibits maximum quantum behavior. 
The Hamiltonian in this scenario takes the form of the quantum Rabi model:
\begin{align}
\hat{H}=&\hbar\tilde{\omega}_\text{c}\hat{c}^\dagger\hat{c}
+g\left(\hat{c}+\hat{c}^\dagger\right) \sum_{n=1}^N \hat{\sigma}_x^{(n)}
-\frac{\hbar\omega_\text{e}}{2}\sum_{n=1}^N \hat{\sigma}_z^{(n)} \nonumber\\
=&\hbar\tilde{\omega}_\text{c}\hat{c}^\dagger\hat{c}
+2g\left(\hat{c}+\hat{c}^\dagger\right)\hat{S}_x-\hbar\omega_\text{e}\hat{S}_z,
\end{align}
where $g$ represents the light-matter coupling strength (assumed real without loss of generality) and $\omega_\text{e}$ denotes the molecular excitation energy.
The operators $\hat{S}_x$ and $\hat{S}_z$ are the collective spin operators: $\hat{S}_{x,z}=\sum_{n=1}^N\hat{\sigma}_{x,z}^{(n)}/2$.
In accordance with the Thomas-Reiche-Kuhn sum rule for electronic transitions~\cite{Sakurai-book}:
\begin{align}
\sum_\nu \frac{|\langle \nu|\hat{p}_{n,k}|\mu\rangle|^2}{\epsilon_\nu^{(n)}-\epsilon_\mu^{(n)}}=\frac{m_\text{e}}{2},
\end{align}
we establish a lower bound for the parameter $\alpha$, defined in Eq.~\eqref{eq: alpha}:
\begin{align}
\hbar\alpha\geq g\sqrt{\frac{2N}{N_\text{e}}}\sqrt{\frac{\tilde{\omega}_\text{c}}{\omega_\text{e}}}.
\end{align}
For the numerical calculations in this study, we set $\hbar\alpha=g\sqrt{2N}$, for which $\hbar\tilde{\omega}_\text{c}=\sqrt{(\hbar\omega_\text{c})^2+4Ng^2}$.

In the quantum Rabi model, the system's Hamiltonian commutes with the total spin operator $\hat{\mathbf{S}}^2=\sum_{\alpha=x,y,z}\hat{S}_\alpha^2$, preserving the quantum number $S$.
If all the molecules are initially in their electronic ground states, $S=N/2$.
The quantum dynamics of the coupled molecule-cavity system are obtained by solving the Schrodinger equation $i\hbar \partial|\psi\rangle/\partial t=\hat{H}|\psi\rangle$ using the basis states $\left\{|S,S_z\rangle\right\}\;(S_z=S,\cdots,-S)$ and the initial condition $|\psi\rangle(t=0)=|S,S\rangle$. 
The matrix elements of $\hat{S}_x$ and $\hat{S}_z$ in this basis are
\begin{align}
\langle S,S_z|\hat{S}_x|S,S_z'\rangle=&\left(\frac{\delta_{S_z,S_z'+1}+\delta_{S_z',S_z+1}}{2}\right)\nonumber\\
&\times\sqrt{S(S+1)-S_zS_z'},
\end{align}
\begin{align}
\langle S,S_z|\hat{S}_z|S,S_z'\rangle=&S_z\delta_{S_z,S_z'}.
\end{align}
To excite the molecules with a short laser pulse, we add the following term to the Hamiltonian:
\begin{align}
\hat{H}_\text{ex}=&\hbar\eta f(t)\sum_{n=1}^N\hat{\Lambda}_1^{(n)}
=\sqrt{2}\hbar\eta f(t)\hat{S}_x,
\end{align}
where $\eta$ represents the excitation amplitude, proportional to the laser intensity, and $f(t)$ is a Gaussian envelop function:
\begin{align}
f(t)=e^{-\left[(t-t_0)/\tau_\text{p}\right]^2}\cos(\omega_\text{p}t).
\end{align}
Here, $t_0$, $\tau_\text{p}$, and $\omega_\text{p}$ are the pulse's center, duration, and frequency, respectively.
In our calculations, the molecular excitation energy is set to $\hbar\omega_\text{e}=2\;\text{eV}$, and the pulse frequency is resonant with the lower polariton energy, which is shifted downward from the bare excitation energy by half the Rabi splitting: $\hbar\omega_\text{p}=\hbar\omega_\text{e}-g\sqrt{N}$.
The pulse has a duration of $\tau_\text{p}=3\;\text{fs}$, center $t_0=5\tau_\text{p}$, and amplitude $\hbar\eta=0.3\;\text{eV}$, corresponding to a laser pulse energy of $1\;\text{nJ}$ focused on an area of $10\;\mu\text{m}^2$~\cite{Engel07}.

The cavity field is initially in the vacuum state $|0\rangle$, where $\hat{a}|0\rangle=0$. 
Under the Bogoliubov transformation, this vacuum state becomes a squeezed state
\begin{align}
|r\rangle=\hat{S}(r)|0\rangle,
\end{align}
where the squeezing operator is
\begin{align}
\hat{S}(r)=e^{-r\left(\hat{a}^{\dagger 2}-\hat{a}^2\right)/2}.
\end{align}
It transforms the field operator as $\hat{a}=\hat{S}(r)\hat{c}\hat{S}^\dagger(r)$.
The squeezed state can be expanded in the Fock state basis as~\cite{Agarwal-book}
\begin{align}
|r\rangle=-\frac{1}{\sqrt{\cosh r}} \sum_{n=0}^\infty 
(\tanh r)^n \frac{\sqrt{(2n)!}}{n!2^n}|2n\rangle.
\end{align}
The Wigner function of the squeezed state is
\begin{align}
W(c)=&\frac{2}{\pi}e^{-2|(\cosh r)c+(\sinh r)c^*|^2}\nonumber\\
=&\frac{2}{\pi}\exp\left\{-2\left[e^{2r}(\text{Re}a)^2+e^{-2r}(\text{Im}a)^2\right]\right\},
\end{align}
where the variance of the cavity field is scaled by $e^{\pm 2r}$ along the real (imaginary) axis.
The cavity photon number operator is
\begin{align}
\hat{n}_\text{c}=&\hat{a}^\dagger\hat{a}\nonumber\\
=&(\cosh 2r)\hat{c}^\dagger\hat{c}+(\sinh r)^2-\frac{\sinh 2r}{2}\left(\hat{c}^\dagger\hat{c}^\dagger+\hat{c}\hat{c}\right)
\end{align}
with the corresponding Weyl symbol
\begin{align}
\left(\hat{n}_\text{c}\right)_\text{W}=&
\cosh 2r\left(|c|^2-\frac{1}{2}\right)+(\sinh r)^2 \nonumber\\
&-\frac{\sinh 2r}{2}\left(c^{*2}+c^2\right).
\end{align}
Under TWA, the equations of motion for the classical variables are
\begin{align}
\frac{\text{d}c}{\text{d}t}=&-i\tilde{\omega}_\text{c}c-\frac{i\sqrt{2}g}{\hbar}\sum_{n=1}^N\Lambda_1^n,
\label{eq: TWA single spin 1}
\end{align}
\begin{align}
\frac{\text{d}\Lambda_1^n}{\text{d}t}=&\omega_\text{e}\Lambda_2^n,
\end{align}
\begin{align}
\frac{\text{d}\Lambda_2^n}{\text{d}t}=&-\omega_\text{e}\Lambda_1^n
-\left[\frac{4g\text{Re}c}{\hbar}+\sqrt{2}\eta f(t)\right]\Lambda_3^n,
\end{align}
\begin{align}
\frac{\text{d}\Lambda_3^n}{\text{d}t}=&\left[\frac{4g\text{Re}c}{\hbar}+\sqrt{2}\eta f(t)\right]\Lambda_2^n.
\label{eq: TWA single spin 4}
\end{align}

Alternatively, when the total spin $S=N/2$ is sufficiently large, the Holstein-Primakoff transformation~\cite{Holstein40} and a Taylor expansion can be employed to express the collective spin operators in terms of a bosonic operator:
\begin{align}
\hat{S}_z=&S-\hat{b}^\dagger\hat{b},\\
\hat{S}_+=&\sqrt{2S-\hat{b}^\dagger\hat{b}}\hat{b}\simeq \sqrt{2S}\left(1-\frac{\hat{b}^\dagger\hat{b}}{4S}\right)\hat{b},\\
\hat{S}_-=&\hat{b}^\dagger \sqrt{2S-\hat{b}^\dagger\hat{b}}\simeq \sqrt{2S}\hat{b}^\dagger\left(1-\frac{\hat{b}^\dagger\hat{b}}{4S}\right),\\
\hat{S}_x=&\frac{\hat{S}_+ + \hat{S}_-}{2}, \hat{S}_y=\frac{\hat{S}_+ - \hat{S}_-}{2i}.
\end{align}
This allows us to rewrite the total Hamiltonian as 
\begin{align}
\hat{H}=&\hbar\tilde{\omega}_\text{c}\hat{c}^\dagger\hat{c}
+\hbar\omega_\text{e}\hat{b}^\dagger\hat{b}
+\sqrt{N}\left[g\left(\hat{c}+\hat{c}^\dagger\right)+\frac{\hbar\eta}{\sqrt{2}}f(t)\right]\nonumber\\
&\times\left[\left(1-\frac{\hat{b}^\dagger\hat{b}}{2N}\right)\hat{b}+\hat{b}^\dagger\left(1-\frac{\hat{b}^\dagger\hat{b}}{2N}\right)\right],
\end{align}
neglecting constants.
When applying TWA to both the bosonic operators $\hat{b}$ and $\hat{c}$, the corresponding equations of motion for the classical variables are
\begin{align}
\frac{\text{d}c}{\text{d}t}=&-i\tilde{\omega}_\text{c}c
-\frac{2ig\sqrt{N}\text{Re}b}{\hbar}\left[1-\frac{|b|^2-1}{2N}\right],
\label{eq: TWA Holstein-Primakoff 1}
\end{align}
\begin{align}
\frac{\text{d}b}{\text{d}t}=&-i\omega_\text{e}b
+i\sqrt{N}\left[\frac{2g\text{Re}c}{\hbar}+\frac{\eta f(t)}{\sqrt{2}}\right]\nonumber\\
&\times\left[-1+\frac{b^2+2|b|^2-1}{2N}\right].
\label{eq: TWA Holstein-Primakoff 2}
\end{align}

\begin{figure}[tbp] 
  \centering
  \includegraphics[width=3.4in, keepaspectratio]{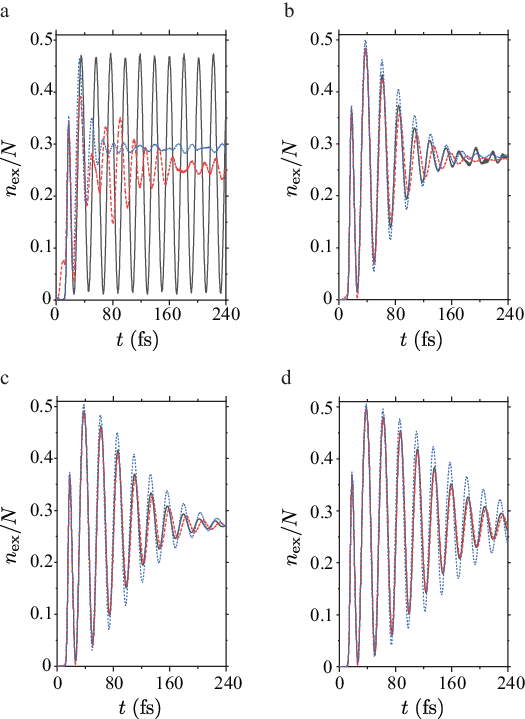}
  \caption{Time evolution of the molecular excitation number $n_\text{ex}$ obtained from quantum dynamic simulations (black solid line), compared to results from applying the truncated Wigner approximation (TWA) to either the single spin operators $\hat{\Lambda}_{i=1,2,3}^{(n)}$ (red dashed line) or the bosonic operator $\hat{b}$ via the Holstein-Primakoff transformation (blue dotted line). 
Results are shown for systems with (a) $N=1$, (b) $N=8$, (c) $N=16$, and (d) $N=32$ molecules. 
The collective light-matter coupling strength is set to $g\sqrt{N}=0.1\;\text{eV}$.}
  \label{fig: compare dynamics of molecular excitation for coupling strength 0.1 eV}
\end{figure}

\begin{figure}[tbp] 
  \centering
  \includegraphics[width=3.4in, keepaspectratio]{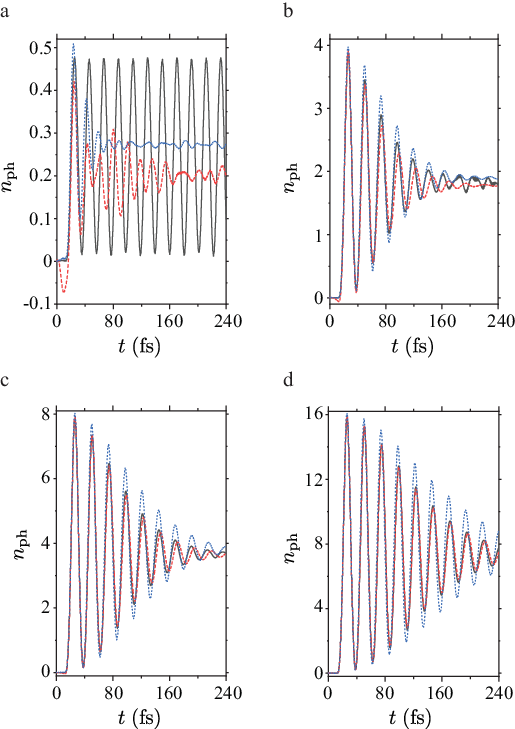}
  \caption{Time evolution of the cavity photon number $n_\text{ph}$ obtained from quantum dynamic simulations (black solid line), compared to results from applying TWA to either the single spin operators $\hat{\Lambda}_{i=1,2,3}^{(n)}$ (red dashed line) or the bosonic operator $\hat{b}$ using the Holstein-Primakoff transformation (blue dotted line). 
Data are presented for systems with (a) $N=1$, (b) $N=8$, (c) $N=16$, and (d) $N=32$ molecules. 
The collective light-matter coupling strength is $g\sqrt{N}=0.1\;\text{eV}$.}
  \label{fig: compare dynamics of cavity photon number for coupling strength 0.1 eV}
\end{figure}

Figures~\ref{fig: compare dynamics of molecular excitation for coupling strength 0.1 eV} and ~\ref{fig: compare dynamics of cavity photon number for coupling strength 0.1 eV} compare the time evolution of the number of molecular excitations, $n_\text{ex}=\langle N/2-\hat{S}_z\rangle$, and the cavity photon number, $n_\text{ph}=\langle \hat{n}_\text{c}\rangle$, as obtained from quantum dynamic simulations with the results from applying the TWA to either the single-spin operators $\hat{\Lambda}_{i=1,2,3}^{(n)}$ (Eqs.~\eqref{eq: TWA single spin 1}--\eqref{eq: TWA single spin 4}) or the bosonic operator $\hat{b}$ via the Holstein-Primakoff transformation (Eqs.~\eqref{eq: TWA Holstein-Primakoff 1}--\eqref{eq: TWA Holstein-Primakoff 2}). 
These comparisons are shown for systems with varying numbers of molecules, $N$, while keeping the collective light-matter coupling strength constant at $g\sqrt{N}=0.1\;\text{eV}=0.05\hbar\omega_\text{e}$ and the resonance condition $\omega_\text{c}=\omega_\text{e}$. 
As $N$ increases, the agreement between the semiclassical TWA results and the exact quantum simulations improves significantly. 
This trend is evident in the time evolution of both the molecular excitation number and the cavity photon number.  
For small systems (e.g., $N=1$), the agreement is poor due to the high nonlinearity associated with the two-level electronic states of a single molecule. 
However, as the system size increases, the enhanced mean-field nature and the reduction in quantum correlation and nonlinear effects leads to a remarkable improvement in the agreement between the TWA and quantum dynamics. 
For $N=8$, the TWA already produces fairly accurate results, and by $N=32$, the agreement with the quantum simulations is excellent, particularly when using the single-spin operators $\hat{\Lambda}_{i=1,2,3}^{(n)}$. 
These results highlight the increasing mean-field character of the system as the number of molecules grows. 
Notably, applying TWA to the single-spin operators yields better results than applying TWA to the bosonic operator $\hat{b}$, especially for larger systems.
For small $N$, the TWA approach using $\hat{\Lambda}_{i=1,2,3}^{(n)}$ exhibits some initial unphysical behavior in the time evolution of the cavity photon number, where negative values appear briefly before the excitation pulse is applied. 
However, this issue is quickly resolved as the number of molecules increases, and it becomes negligible in systems with larger $N$. 

\begin{figure}[tbp] 
  \centering
  \includegraphics[width=3.4in, keepaspectratio]{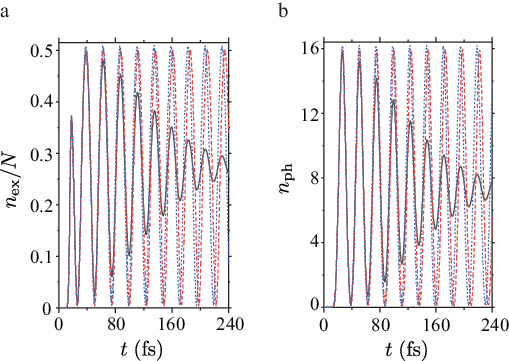}
  \caption{Time evolution of (a) molecular excitation number $n_\text{ex}$ and (b) cavity photon number $n_\text{ph}$, comparing quantum dynamic simulations (black solid line) with pure mean-field theory results applied to either the single-spin operators $\hat{\Lambda}_{i=1,2,3}^{(n)}$ (red dashed line) or the bosonic operator $\hat{b}$ (blue dotted line).}
  \label{fig: compare the mean-field result with quantum dynamic simulation}
\end{figure}

Although the increasing accuracy of the TWA results in larger systems can be attributed to the enhanced mean-field behavior, it is important to note that pure mean-field calculations (i.e., without sampling from the Wigner distribution) yield significantly less accurate predictions. 
Figure~\ref{fig: compare the mean-field result with quantum dynamic simulation} illustrates this by comparing the time evolution of both the molecular excitation number and the cavity photon number, as obtained from quantum dynamic simulations, with those from mean-field theory applied to either the single-spin operators $\hat{\Lambda}_{i=1,2,3}^{(n)}$ or the bosonic operator $\hat{b}$. 
In the mean-field approach, all physical observables in the classical equations of motion (Eqs.~\eqref{eq: TWA single spin 1}--\eqref{eq: TWA single spin 4} and \eqref{eq: TWA Holstein-Primakoff 1}--\eqref{eq: TWA Holstein-Primakoff 1}) are replaced by their average values, which are assumed to be the same for all molecules since the molecules are equivalent. 
The equations of motions for the average values in $\hat{\Lambda}_{i=1,2,3}^{(n)}$-mean-field theory are given by
\begin{align}
\frac{\text{d}\bar{c}}{\text{d}t}=&-i\tilde{\omega}_\text{c}\bar{c}-\frac{i\sqrt{2}gN}{\hbar}\bar{\Lambda}_1,
\end{align}
\begin{align}
\frac{\text{d}\bar{\Lambda}_1}{\text{d}t}=&\omega_\text{e}\bar{\Lambda}_2,
\end{align}
\begin{align}
\frac{\text{d}\bar{\Lambda}_2}{\text{d}t}=&-\omega_\text{e}\bar{\Lambda}_1
-\left[\frac{4g\text{Re}\bar{c}}{\hbar}+\sqrt{2}\eta f(t)\right]\bar{\Lambda}_3,
\end{align}
\begin{align}
\frac{\text{d}\bar{\Lambda}_3}{\text{d}t}=&\left[\frac{4g\text{Re}\bar{c}}{\hbar}+\sqrt{2}\eta f(t)\right]\bar{\Lambda}_2.
\end{align}
The initial values of the averages are $\bar{c}(t=0)=0$, $\bar{\Lambda}_1(t=0)=\bar{\Lambda}_2(t=0)=0$, and $\bar{\Lambda}_3(t=0)=1/\sqrt{2}$.
Similarly, the equations of motions for the average values in $\hat{b}$-mean-field theory are given by
\begin{align}
\frac{\text{d}\bar{c}}{\text{d}t}=&-i\tilde{\omega}_\text{c}\bar{c}
-\frac{2ig\sqrt{N}\text{Re}\bar{b}}{\hbar}\left[1-\frac{|\bar{b}|^2-1}{2N}\right],
\end{align}
\begin{align}
\frac{\text{d}\bar{b}}{\text{d}t}=&-i\omega_\text{e}\bar{b}
+i\sqrt{N}\left[\frac{2g\text{Re}\bar{c}}{\hbar}+\frac{\eta f(t)}{\sqrt{2}}\right]\nonumber\\
&\times\left[-1+\frac{\bar{b}^2+2|\bar{b}|^2-1}{2N}\right].
\end{align}
The initial condition is $\bar{c}(t=0)=0,\bar{b}(t=0)=0$.

\begin{figure}[tbp] 
  \centering
  \includegraphics[width=3.4in, keepaspectratio]{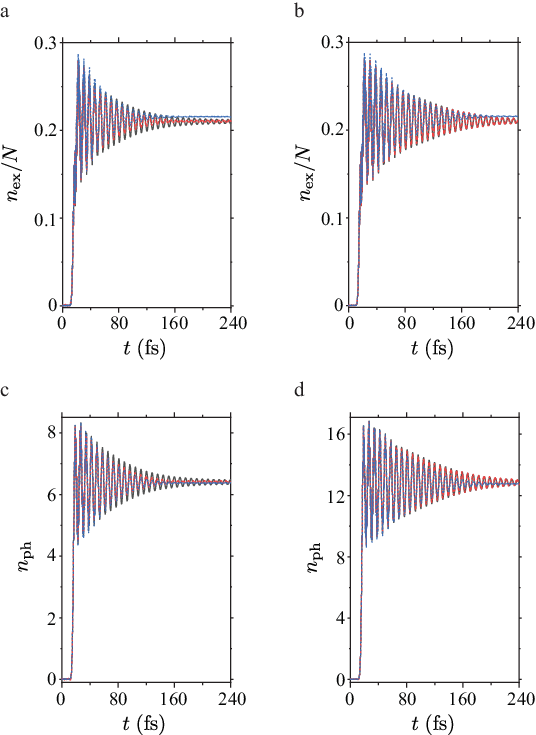}
  \caption{Time evolution of (a, b) molecular excitation number $n_\text{ex}$ and (c, d) cavity photon number $n_\text{ph}$ for systems of (a, c) $N=32$ and (b, d) $N=64$ molecules. 
Comparisons are shown between quantum dynamic simulations (black solid line) and results obtained from applying TWA to either the single-spin operators $\hat{\Lambda}_{i=1,2,3}^{(n)}$ (red dashed line) or the bosonic operator $\hat{b}$ (blue dotted line). 
The collective light-matter coupling strength is $g\sqrt{N}=0.3\;\text{eV}$.}
  \label{fig: compare TWA with quantum dynamics for coupling strength 0.3 eV}
\end{figure}

For stronger collective light-matter coupling ($g\sqrt{N}=0.3\;\text{eV}=0.15\hbar\omega_\text{e}$), which approaches the ultrastrong coupling regime~\cite{Kockum19}, Fig.~\ref{fig: compare TWA with quantum dynamics for coupling strength 0.3 eV} shows the time evolution of the molecular excitation and cavity photon numbers for systems with $N=32$ and $N=64$ molecules. 
While the agreement between the semiclassical TWA results and exact quantum dynamics is slightly reduced compared to the weaker coupling case ($g\sqrt{N}=0.1\;\text{eV}$), the TWA still provides reasonably accurate predictions. 
Importantly, as with the weaker coupling case, increasing the number of molecules further improves the agreement between the TWA and quantum simulations. 

Next, we consider systems where nuclear degrees of freedom are included, introducing vibronic coupling between the electronic and nuclear states. 
The potential energy surface of the electronic ground state is modeled by a harmonic potential $V_\text{g}(q)=M\omega_\text{v}^2q^2/2$, where $M$, $\omega_\text{v}$, and $q$ represent the effective mass, frequency, and coordinate of a single vibrational mode.
The potential energy surface of the electronic excited state is similarly modeled but displaced by $q_0$ to represent vibronic coupling:
\begin{align}
V_\text{e}(q)=\hbar\omega_\text{e}+\frac{M\omega_\text{v}^2(q-q_0)^2}{2}.
\end{align}
The reorganization energy $\lambda_0$, which quantifies the vibronic coupling strength, is related to the displacement $q_0$ by $\lambda_0=M\omega_\text{v}^2q_0^2/2$, and the Frank-Condon excitation energy is $\Delta\epsilon=\hbar\omega_\text{e}+\lambda_0$. 
For simplicity, we assume that the transition dipole moment is independent of the nuclear coordinate. 

In the presence of vibronic coupling, the system's total Hamiltonian no longer commutes with the collective spin operator $\hat{\mathbf{S}}^2$, meaning that the quantum number $S$ is no longer conserved.
Thus, to account for this, we apply the TWA to the single-spin operators $\hat{\Lambda}_{i=1,2,3}^{(n)}$. 
The corresponding equations of motion for the classical variables are derived from the TWA and now include terms for the nuclear degrees of freedom. 
These equations describe the time evolution of the photonic, electronic, and nuclear variables, while accounting for the vibronic coupling:
\begin{align}
\frac{\text{d}c}{\text{d}t}=-i\tilde{\omega}_\text{c}c-\frac{i\sqrt{2}g}{\hbar}\sum_{n=1}^N \Lambda_1^n,
\end{align}
\begin{align}
\frac{\text{d}\Lambda_1^n}{\text{d}t}=\left[\omega_\text{e}-\frac{M\omega_\text{v}^2q_0}{\hbar}\left(q_n-\frac{q_0}{2}\right)\right]\Lambda_2^n,
\end{align}
\begin{align}
\frac{\text{d}\Lambda_2^n}{\text{d}t}=&\left[-\omega_\text{e}+\frac{M\omega_\text{v}^2q_0}{\hbar}\left(q_n-\frac{q_0}{2}\right)\right]\Lambda_1^n \nonumber\\
&-\left[\frac{4g\text{Re}c}{\hbar}+\sqrt{2}\eta f(t)\right]\Lambda_3^n,
\end{align}
\begin{align}
\frac{\text{d}\Lambda_3^n}{\text{d}t}=\left[\frac{4g\text{Re}c}{\hbar}+\sqrt{2}\eta f(t)\right]\Lambda_2^n,
\end{align}
\begin{align}
\frac{\text{d}q_n}{\text{d}t}=\frac{p_n}{M},
\end{align}
\begin{align}
\frac{\text{d}p_n}{\text{d}t}=-M\omega_\text{v}^2\left(q_n-\frac{q_0}{2}\right)
-\frac{M\omega_\text{v}^2q_0}{\sqrt{2}}\Lambda_3^n,
\end{align}
where $q_n$ and $p_n$ are the nuclear coordinate and momentum of the $n$-th molecule. 
It is assumed that there is no initial correlation between the nuclear degrees of freedom of different molecules so that the nuclear Wigner function of the total system is given by the product $W_\text{nu}=\prod_{n=1}^N W_\text{nu}^{(n)}$.
The initial state of the nuclear degrees of freedom is assumed to be the vibrational ground state of the electronic ground-state potential energy surface $V_\text{g}(q)$, with the nuclear Wigner function of each molecule given by a Gaussian distribution:
\begin{align}
W_\text{nu}^{(n)}(q_n,p_n)=\frac{1}{\pi\sigma_q\sigma_p}\exp\left[-\left(\frac{q_n^2}{\sigma_q^2}+\frac{p_n^2}{\sigma_p^2}\right)\right],
\end{align}
where $\sigma_q=\sqrt{\hbar/(M\omega_\text{v})}$ and $\sigma_p=\sqrt{\hbar M\omega_\text{v}}$.
In the following numerical calculations, the vibrational frequency is set to $\hbar\omega_\text{v}=0.1\;\text{eV}$. 
The cavity frequency is taken to be in resonance with the Frank-Condon excitation energy $\hbar\omega_\text{c}=\Delta\epsilon$, and the laser pulse frequency is set to $\hbar\omega_\text{p}=\Delta\epsilon-g\sqrt{N}$, accounting for the Rabi splitting.

To investigate the validity of the TWA approach for molecular exciton-polariton dynamics, we focus on the quantum coherence between electronic excitations in different molecules. 
According to polaron decoupling effect~\cite{Spano15, Herrera16, Phuc19, Takahashi20, Phuc21}, strong light-matter coupling can suppress the decay of quantum coherence caused by interactions with nuclear degrees of freedom.
The quantum coherence operator, averaged over all pairs of molecules, is defined as
\begin{align}
\hat{C}_\text{ee}=&\frac{1}{N(N-1)}\sum_{n=1}^N\sum_{m\not=n}
|\text{e}_n\rangle\langle\text{g}_n|\otimes|\text{g}_m\rangle\langle\text{e}_m|\nonumber\\
=&\frac{1}{2N(N-1)}\sum_{n=1}^N\sum_{m\not=n}
\left(\hat{\Lambda}_1^n-i\hat{\Lambda}_2^n\right)\left(\hat{\Lambda}_1^m+i\hat{\Lambda}_2^m\right).
\end{align}
Under the TWA, the expectation value of the quantum coherence $\mathcal{C}_\text{ee}=\langle \hat{C}_\text{ee}\rangle$ is approximated by averaging the classical variable
\begin{align}
C_\text{ee}=\frac{1}{2N(N-1)}\sum_{n=1}^N\sum_{m\not=n}
\left(\Lambda_1^n-i\Lambda_2^n\right)\left(\Lambda_1^m+i\Lambda_2^m\right)
\end{align}
since the single-spin operators for different molecules commute.
Moreover, $\mathcal{C}_\text{ee}$ is a real number, as the correlation function changes to its complex conjugate under the exchange of two molecules.

\begin{figure}[tbp] 
  \centering
 \includegraphics[width=3.4in, keepaspectratio]{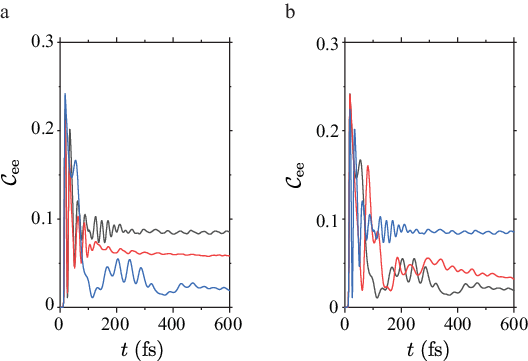}
  \caption{Time evolution of the quantum coherence $\mathcal{C}_\text{ee}$ between electronic excitations of different molecules for a system with $N=32$ molecules. (a) The collective coupling strength is fixed at $g\sqrt{N}=0.1\;\text{eV}$, while the reorganization energy is varied: $\lambda_0=0.01\;\text{eV}$ (black solid line), $\lambda_0=0.03\;\text{eV}$ (red dashed line), and $\lambda_0=0.1\;\text{eV}$ (blue dotted line). (b) The reorganization energy is fixed at $\lambda_0=0.01\;\text{eV}$, while the collective coupling strength is varied: $g\sqrt{N}=0.01\;\text{eV}$ (black solid line), $g\sqrt{N}=0.03\;\text{eV}$ (red dashed line), and $g\sqrt{N}=0.1\;\text{eV}$ (blue dotted line).}
  \label{fig: compare quantum coherence}
\end{figure}

Figure~\ref{fig: compare quantum coherence} shows the time evolution of the quantum coherence $\mathcal{C}_\text{ee}$ for a system of $N=32$ molecules under various reorganization energies and collective coupling strengths. 
The steady-state value of $\mathcal{C}_\text{ee}$ is found to increase either with decreasing reorganization energy (when the coupling strength is held constant) or with increasing light-matter coupling (when the reorganization energy is fixed). 
This behavior is consistent with the polaron decoupling effect, where strong light-matter coupling mitigates the decoherence caused by nuclear vibrational interactions.

\section{Conclusion}
We have developed a semiclassical theory based on the truncated Wigner approximation (TWA) to study the dynamics of molecular exciton polaritons under strong light-matter coupling. 
Initially, we applied the TWA to a simplified system of two-level molecules (spin-1/2 systems) without vibronic coupling, focusing on the purely electronic degrees of freedom interacting with an optical cavity mode. 
The results obtained from this model were validated by comparing them with exact quantum dynamic simulations. 
Notably, the TWA demonstrated excellent agreement with quantum simulations for systems containing a large number of molecules, even in the ultrastrong coupling regime. 
This is crucial, as large molecular ensembles are typically encountered in experimental setups involving optical cavities, and collective nature of exciton polaritons becomes more pronounced in such systems. 

One of the key findings from our study is the increasing accuracy of the TWA as the system size grows. 
This can be attributed to the enhanced mean-field behavior in larger systems, where quantum correlations and nonlinear effects become less prominent. 
The agreement between the TWA and full quantum dynamics highlights the capability of the semiclassical approach to capture the essential physics of molecular exciton polaritons, including the intricate balance between collective excitations and light-matter interactions. 

We next extended the TWA framework to include nuclear degrees of freedom, incorporating vibronic coupling into the dynamics. 
This extension allowed us to examine how strong light-matter coupling influences the interaction between electronic and nuclear motions in molecular systems. 
The vibronic coupling introduces decoherence and dissipation into the system, factors that are critical in determining the overall behavior of molecular excitations. 
Our results revealed that strong light-matter coupling can mitigate the decoherence effects typically caused by nucelar vibrations, preserving quantum coherence between molecular excitations over extended periods. 
This phenomenon, known as the dynamic polaron decoupling effect, has profound implications for the stability of quantum states in molecular systems and could lead to enhanced control over photochemical reactions in polaritonic environments. 

Our work laid the groundwork for future research into more complex molecular systems and light-matter interactions. 
The TWA framework developed here can be applied to multi-level molecular systems, where the interactions between multiple excited states and cavity modes may lead to even richer dynamics. 
Future applications of this approach could provide deeper insights into the design of polaritonic materials, the control of photochemical reactions, and the development of new technologies based on light-matter interactions in confined environments. 
  

\begin{acknowledgements}
N. T. Phuc acknowledges financial support from Hirose Foundation.
The computations were performed using Research Center for Computational Science, Okazaki, Japan.
\end{acknowledgements}





\end{document}